\documentclass[oneside, a4paper]{article}
\usepackage{authblk}
\usepackage{amsmath,graphicx,amssymb,amsfonts,mathabx}
\usepackage{subcaption, xcolor}
\usepackage{algpseudocode,algorithm,algorithmicx}

\makeatletter
\def\widebreve{\mathpalette\wide@breve}
\def\wide@breve#1#2{\sbox\z@{$#1#2$}%
     \mathop{\vbox{\m@th\ialign{##\crcr
\kern0.08em\brevefill#1{0.8\wd\z@}\crcr\noalign{\nointerlineskip}%
                    $\hss#1#2\hss$\crcr}}}\limits}
\def\brevefill#1#2{$\m@th\sbox\tw@{$#1($}%
  \hss\resizebox{#2}{\wd\tw@}{\rotatebox[origin=c]{90}{\upshape(}}\hss$}
\makeatletter

\newcommand\bs[1]{\boldsymbol{#1}}
\newcommand\mb[1]{\mathbf{#1}}
\newcommand\mc[1]{\mathcal{#1}}
\newcommand\bb[1]{\mathbb{#1}}
\newcommand\wt[1]{\widetilde{#1}}
\newcommand\wb[1]{\overline{#1}}
\newcommand\wh[1]{\widehat{#1}}
\newcommand\wc[1]{\widecheck{#1}}
\newcommand\wbr[1]{\widebreve{#1}}

\algrenewcommand\algorithmicrequire{\textbf{Note:}}
\algrenewcommand\algorithmicensure{\textbf{Note:}}

\newcommand{\norm}[1]{\left\lVert#1\right\rVert}
\DeclareMathOperator{\vect}{vec}
\DeclareMathOperator{\vecth}{vech}

\title{Relative Kinematics Estimation Using Accelerometer Measurements}

\author[\empty]{Anurodh Mishra \thanks{This work is partially funded by the European Leadership Joint Undertaking (ECSEL JU), under grant agreement No 876019, the ADACORSA project - ”Airborne Data Collection on Resilient System Architectures.”}}
\author[\empty]{Raj Thilak Rajan}
\affil[\empty]{Circuits \& Systems Group, Department of Microelectronics, Faculty of EEMCS, Delft  University of Technology, Delft, The Netherlands}
\date{}                     
\begin{document}
\maketitle
\begin{abstract} 
Given a network of $N$ static nodes in $D$-dimensional space and the pairwise distances between them, the challenge of estimating the coordinates of the nodes is a well-studied problem. However, for numerous application domains, the nodes are mobile and the estimation of relative kinematics (e.g., position, velocity and acceleration) is a challenge, which has received limited attention in literature. In this paper, we introduce a time-varying Grammian-based data model for estimating the relative kinematics of mobile nodes with polynomial trajectories, given the time-varying pairwise distance measurements between the nodes. Furthermore, we consider a scenario where the nodes have on-board accelerometers, and extend the proposed data model to include these accelerometer measurements. We propose closed-form solutions to estimate the relative kinematics, based on the proposed data models. We conduct simulations to showcase the performance of the proposed estimators,  which show improvement against state-of-the-art methods.
\end{abstract}

\section{Introduction} \label{sec:intro}

The problem of estimating the position coordinates of $N$ points, in $D$-dimensional space, given a \textit{dissimilarity} measure, has a long history in scientific literature \cite{Tongerson1952,Gower1982,Gower1985,Hayden1990}. If these dissimilarities are represented by Euclidean Distance Matrices (EDMs), then Multidimensional scaling (MDS) can be employed to estimate the relative positions of the $N$ points. Given the pairwise distances between nodes, various estimators have been proposed for the relative localization of the nodes in a network \cite{Alfakih1999,Biswas2004,Dokmanic2015}. However, in numerous applications involving motion systems, such as robot swarms \cite{Cornejo2015}, the nodes are mobile and measurements of pairwise distances between these nodes are available over time. In such cases, it is useful to model this time dependency in order to understand the underlying relative kinematics of the nodes, particularly in networks where position references (or anchors) are unavailable.

To the best of our knowledge, the earliest work on time-varying Euclidean distance measurements was proposed in \cite{Rajan2013,Rajan2019}, where the authors presented a systematic way of estimating higher-order relative kinematics for a network of mobile nodes from time-varying distance measurements, where each node has a polynomial trajectory in time. However, to uniquely estimate the relative kinematics, additional rigid-body constraints are required. More recently, a Grammian-based approach for recovering trajectories from time-varying pairwise distances was proposed \cite{Tabaghi2020}, using spectral factorization methods. However, the proposed solutions require anchor measurements.


In this paper, we aim to estimate the relative kinematics of a network of mobile nodes given the time-varying pairwise distances measurements without any apriori knowledge of anchor nodes or references in the network. The main advantage of the proposed algorithm over the state-of-the-art in \cite{Rajan2019} is that it does not require additional rigid body constraints to be solved uniquely. To this end, in Section \ref{sec:data_model_no_acc}, we propose an alternative formulation to the data model presented in \cite{Rajan2019}. In Section \ref{sec:data_model_acc}, we modify the derived data model to incorporate accelerometer measurements under certain assumptions. We conduct simulations and present the results in Section \ref{sec:sim}, which show the benefits of the proposed solutions.


\textit{Notation:} Lower case alphabets, e.g., $a$, represents scalars and bold-faced lower case letters, e.g., $\mb{a}$, denote a column vector. A bold capital letter, e.g., $\mb{A}$, indicates a matrix and calligraphic letters e.g., $\mc{A}(\cdot)$ represent matrices that are explicitly shown to be a function of a vector or another matrix. Half-vectorization of a symmetric matrix $\mb{A}$ is denoted by $\vecth(\mb{A})$, and a simple vectorization is represented by $\vect(\mb{A})$. The symbol $\otimes$ denotes a Kronecker product. A vector and matrix of real-valued entries are denoted by $\bb{R}^{N}$ and $\bb{R}^{M \times N}$, respectively. A column vector of ones with length $N$ is denoted by $\mb{1}_N$, and the $l_2$-norm is denoted by $\norm{\cdot}$. Given a positive semidefinite matrix, $\mb{D} \in \bb{R}^{N \times N}$, constructed using an underlying point set $\mb{X} \in \bb{R}^{D \times N}$, an estimate of the point set using classical Multidimensional scaling (MDS), is given by \begin{align}
        \bs{\mc{F}}_{\text{mds}}\left(\mb{D}, \mb{X}\right) &\triangleq\ \arg \min_{\mb{X}} \: \norm{ \mb{D} - \mb{X}^T \: \mb{X}} \: \: \text{s.t.} \: \text{rank}(\mb{X}) = D \nonumber \\ 
        &= \bs{\Lambda}^{1/2} \: \mb{V}^T
\end{align} where $\bs{\Lambda}$ contains the first $D$ non-zero Eigenvalues of $\mb{D}$, and $\mb{V}$ contains the corresponding Eigenvectors \cite{Borg2005}.


\section{Preliminaries} \label{sec:prelim}

Consider a system of $N$ mobile nodes in $D$-dimensional Euclidean space, whose trajectory can be modelled as an $L+1$th order polynomial in time $t$ , i.e.,  $\wb{\mb{S}}(t) = \sum_{l = 0}^L \left( l ! \right)^{-1} \: \wb{\mb{Y}}_l \: t^l$ where $\wb{\mb{S}}(t)$ is the polynomial trajectory as a function of time $t$ \cite{Rajan2019}. Furthermore, we define the $l$th order derivative of this polynomial as $\wb{\mb{Y}}_l = \wb{\mb{S}}^{(l)}(t) \vert_{t = 0}$, for $l \in \{0, 1, \ldots , L \}$, which are assumed to be finite. We define the time-varying Euclidean Distance Matrix (EDM) of the network as  \begin{equation}
    \mb{D}(t) \triangleq \mb{1}_N \: \text{diag}(\wb{\mb{G}}(t))^T - 2 \: \wb{\mb{G}}(t) + \text{diag}(\wb{\mb{G}}(t)) \: \mb{1}_N^T \label{eq:edm}
\end{equation} where $\wb{\mb{G}}(t) \triangleq \wb{\mb{S}}(t)^T \: \wb{\mb{S}}(t)$ is the time-varying Grammian. The position coordinates at time instant $t_k$ is given as $\wb{\mb{X}}_k \triangleq \wb{\mb{S}}(t) \vert_{t = t_k}$, and the acceleration
$\ddot{\wb{\mb{X}}}_k$ is obtained by twice differentiating $\wb{\mb{S}}(t)$ w.r.t. time i.e., \begin{equation}
    \ddot{\wb{\mb{X}}}_k = \dfrac{\partial^2 \mb{S}(t)}{\partial t^2} \Big \vert_{t = t_k} = \sum_{l = 2}^L \left( (l-2)! \right)^{-1} \: \wb{\mb{Y}}_l \: t_k^{l - 2}. \label{eq:poly_traj_acc}
\end{equation}



Now, the time-varying position and acceleration coordinates centered at the origin at time $t_k$ is given by
\begin{subequations}
    \begin{align}
        \mb{X}_k &= \sum_{l = 0}^L \left( l ! \right)^{-1} \: \mb{Y}_l \: t_k^l\\
        \ddot{\mb{X}}_k &= \sum_{l = 2}^{L-2} \left( (l - 2) ! \right)^{-1} \: \mb{Y}_l \: t_k^{l - 2}
    \end{align} \label{eq:poly_traj_centered}
\end{subequations} where $\mb{Y}_l \triangleq \wb{\mb{Y}}_l \: \mb{C}$ and $\mb{C} = \mb{I}_N - N^{-1}\mb{1}_N^T\mb{1}_N$ is the centering matrix \cite{Gower1982}. The Grammian for the centered coordinates $\mb{X}_k$ at time $t_k$, denoted by $\mb{G}_k$, can be calculated by double centering the EDM from (\ref{eq:edm}) at time $t_k$, yielding,
\begin{equation}
    \mb{G}_k = -\dfrac{1}{2} \: \mb{C} \: \mb{D}_k \: \mb{C} = \dfrac{1}{2} \: \mb{C} \: \left(2 \: \wb{\mb{X}}_k^T \: \wb{\mb{X}}_k \right) \: \mb{C} = \mb{X}_k^T \: \mb{X}_k  \label{eq:centered_grammian}
\end{equation}where $\mb{D}_k$ denotes the EDM at time instant $t_k$. Using (\ref{eq:poly_traj_centered}) for $\mb{X}_k$, the Grammian, $\mb{G}_k$ (\ref{eq:centered_grammian}), can be rewritten as
\begin{equation}
    \mb{G}_k = \mb{B}_0 + \mb{B}_1 \: t_k + \mb{B}_2 \: t_k^2 + \ldots + \mb{B}_L \: t_k^L \label{eq:time_poly_gramian}    
\end{equation} where \begin{equation}
\label{eq:def_Bl}
    \mb{B}_l = \sum_{m = 0}^l \: \left( m! \: (l - m)! \right)^{-1} \: \mb{Y}_m^T \: \mb{Y}_{l-m}.
\end{equation} Given the distances, $\mb{D}_k$, we aim to estimate $\mb{B}_l$, which subsequently yield the relative kinematics $\mb{Y}_l$ for $l \in \{ 0, 1, \ldots, L \}$. In the following section, we propose algorithms to estimate the relative kinematics, given the distance measurements, which in reality are plagued with noise. 

\section{Pairwise Distances} \label{sec:data_model_no_acc}

\subsection{Data Model with only pairwise distances} 
Vectorizing (\ref{eq:time_poly_gramian}) and using the distributive property of vectorization over summation, we get
\begin{equation}
    \mb{g}_k = \mb{b}_0 + t_k \: \mb{b}_1 + t_k^2 \: \mb{b}_2 + \ldots + t_k^L \: \mb{b}_L \label{eq:time_poly_gram_vec}
\end{equation}
where $\mb{b}_l = \vecth(\mb{B}_l)$, for $l \in \{ 0, 1, \ldots, L \}$ and $\mb{g}_k = \vecth(\mb{G}_k)$. Without loss of generality, let $\wc{\mb{g}}_k = \mb{g}_k + \bs{\eta}_{g_k}$ be the noisy measurement plagued by additive white Gaussian noise with covariance matrix $\bs{\Sigma}_{g_k}$. Stacking the vectorized Grammians $\wc{\mb{g}}_k$ for all $K$ timestamps in column vector $\wc{\mb{g}}$, we get
\begin{equation}
    \mb{T} \: \bs{\theta} = \wc{\mb{g}} \label{eq:lsq_no_acc}
\end{equation}
where $\mb{T} = \begin{bmatrix}
\mb{1}_K \otimes \mb{I}_{\wb{N}}, & \mb{t} \otimes \mb{I}_{\wb{N}}, & \ldots, & \mb{t}^{\odot L} \otimes \mb{I}_{\wb{N}}
\end{bmatrix}$, $\bs{\theta} = \begin{bmatrix}
\mb{b}_0, & \mb{b}_1, & \ldots, & \mb{b}_L
\end{bmatrix}^T$, $\wc{\mb{g}} = \begin{bmatrix}
\wc{\mb{g}}_0, & \wc{\mb{g}}_1, & \ldots, & \wc{\mb{g}}_K 
\end{bmatrix}^T$. Here, $\wb{N} = N(N+1)/2$ and $\mb{t}$ is a column vector of time stamps $t_k$. The unknown $\bs{\theta}$ can then be calculated by solving the following least-squares problem $\arg \min_{\bs{\theta}} \norm{\mb{T} \: \bs{\theta} - \wc{\mb{g}}}_2^2$ leading to a closed-form solution given by
\begin{equation}
    \wh{\bs{\theta}} = \left( \mb{T}^T \: \mb{T} \right)^{-1} \: \mb{T}^T \: \wc{\mb{g}} \label{eq:grammian_coeff_est_no_acc}
\end{equation} which is an optimal estimator given the assumption of additive white Gaussian noise on the measurements. 

\subsection{Relative Kinematics Estimates} \label{sec:no_acc_kin}
Consider a scenario when the nodes are in constant acceleration i.e., $\mb{Y}_l = \mb{0}$ for $l \geq 3$. From (\ref{eq:grammian_coeff_est_no_acc}), the estimates $\wh{\mb{B}}_l$, $l \in \{ 0, 4 \}$ can be reconstructed, and subsequently using (\ref{eq:def_Bl}), the relative position and relative acceleration can be calculated using classical MDS algorithms \cite{Borg2005}, i.e.,
\begin{subequations}
    \begin{align}
        \wh{\mb{Y}}_0 &= \bs{\mc{F}}_{\text{mds}} \left( \wh{\mb{B}}_0, \mb{Y}_0 \right) \label{eq:sol_const_acc_1a}\\
        \wh{\mb{Y}}_2 &= \bs{\mc{F}}_{\text{mds}} \left( 4 \: \wh{\mb{B}}_4, \mb{Y}_2 \right) \label{eq:sol_const_acc_1b}
    \end{align} \label{eq:sol_const_acc_1}
\end{subequations} where $\wh{\mb{Y}}_0$ is the estimate for the centered position coordinates $\mb{Y}_0$ at time $t = 0$ and $\wh{\mb{Y}}_2$ is the estimate of the relative acceleration centered at the origin. Note that the estimates $\wh{\mb{Y}}_0$ and $\wh{\mb{Y}}_2$ from the MDS solution in (\ref{eq:sol_const_acc_1}) are each known only up to a rotation, which we denote by $\mb{H}_0$ and $\mb{H}_2$ respectively. We assume the rotation associated with $\wh{\mb{Y}}_0$ to be identity, i.e. $\mb{H}_0 = \mb{I}_D$. However, we need to estimate the unknown rotation corresponding to $\wh{\mb{Y}}_2$, given by $\mb{H}_2$. Now for $l \in \{1, 3\}$ in (\ref{eq:def_Bl}), $\mb{B}_l$ take the following Lyapunov-like form
\begin{subequations}
    \begin{align}
        \mb{B}_1 &= \mb{Y}_0^T \: \mb{Y}_1 + \mb{Y}_1^T \: \mb{Y}_0\\
        2 \: \mb{B}_3 &= \mb{Y}_2^T \: \mb{Y}_1 + \mb{Y}_1^T \: \mb{Y}_2
        \end{align}
\end{subequations}
Substituting the estimates of $\mb{B}_l$ from (\ref{eq:grammian_coeff_est_no_acc}) for $l \in \{1, 3\}$ and estimates of $\mb{Y}_0$ and $\mb{Y}_2$ from (\ref{eq:sol_const_acc_1}), we get
\begin{subequations}
    \begin{align}
        \wh{\mb{B}}_1 &= \wh{\mb{Y}}_0^T \: \mb{Y}_1 + \mb{Y}_1^T \: \wh{\mb{Y}}_0 \label{eq:lyap_cnst_acc_1a}\\
        2 \: \wh{\mb{B}}_3 &= \wh{\mb{Y}}_2^T \: \mb{H}^T_2  \: \mb{Y}_1 + \mb{Y}_1^T \: \mb{H}_2 \: \wh{\mb{Y}}_2 \label{eq:lyap_cnst_acc_1b}
        \end{align} \label{eq:lyap_cnst_acc_1}
\end{subequations}
where $\mb{H}_2$ is the unknown rotation and $\mb{Y}_1$ is the unknown relative velocity to be estimated. Note that the individual Lyapunov-like equations in (\ref{eq:lyap_cnst_acc_1}) are under-determined and require additional constraints to obtain a unique solution \cite{Rajan2019,Chu1989}. As one of the contributions of this paper, we propose a solution to the combined set of equations in (\ref{eq:lyap_cnst_acc_1}) for estimating $\mb{Y}_1$ and $\mb{H}_2$, as opposed to the approach in \cite{Rajan2019}.We begin by rewriting (\ref{eq:lyap_cnst_acc_1}),
\begin{subequations}
    \begin{align}
        \wbr{\mb{B}}_1 &=\begin{bmatrix}
        \bs{\Lambda}_0 & \mb{0}
        \end{bmatrix}^T \: \mb{Z} + \mb{Z}^T \: \begin{bmatrix}
        \bs{\Lambda}_0 & \mb{0}
        \end{bmatrix} \\
        \wbr{\mb{B}}_3 &= \begin{bmatrix}
        \bs{\Lambda}_2 & \mb{0}
        \end{bmatrix}^T \: \wb{\mb{Z}} + \wb{\mb{Z}}^T \: \begin{bmatrix}
        \bs{\Lambda}_2 & \mb{0}
        \end{bmatrix}
    \end{align} \label{eq:lyap_chu_form}
\end{subequations}
where $\mb{Z} \triangleq \begin{bmatrix}
    \mb{Z}_1 & \mb{Z}_2
\end{bmatrix} = \mb{U}^T_0 \: \mb{Y}_1 \: \mb{V}_0$, $\wb{\mb{Z}} \triangleq \begin{bmatrix}
    \wb{\mb{Z}}_1 & \wb{\mb{Z}}_2
\end{bmatrix} = \mb{U}^T_2 \: \left( \mb{H}_2^T \: \mb{Y}_1 \right) \: \mb{V}_2$, $\wbr{\mb{B}}_1 = \mb{V}^T_0 \: \wh{\mb{B}}_1 \: \mb{V}_0$ and $\wbr{\mb{B}}_3 = \mb{V}^T_2 \: \wh{\mb{B}}_3 \: \mb{V}_2$ \cite{Chu1989}. Here $\mb{Z}_1, \wb{\mb{Z}}_1 \in \bb{R}^{D \times D}$ and $\mb{Z}_2, \wb{\mb{Z}}_2 \in \bb{R}^{(N-D) \times D}$. Furthermore, $\mb{U}_0 \in \bb{R}^{D \times D}$, $\mb{V}_0 \in \bb{R}^{N \times N}$  and $\bs{\Lambda}_0 \in \bb{R}^{D \times D}$ are the respective singular vectors and singular values of $\wh{\mb{Y}}_0$. $\mb{U}_2 \in \bb{R}^{D \times D}$, $\mb{V}_2 \in \bb{R}^{N \times N}$  and $\bs{\Lambda}_2 \in \bb{R}^{D \times D}$ are similarly defined for $\wh{\mb{Y}}_2$. Here, $\mb{Z}_2$ and $\wb{\mb{Z}}_2$ can be uniquely determined, while the $D^2 - D$ off-diagonal elements of $\mb{Z}_1$ and $\wb{\mb{Z}}_1$ are unknown \cite{Chu1989}. We introduce
\begin{subequations}
    \begin{align}
        \mb{z} \triangleq \vect(\mb{Z}) &= \mb{K}_0 \: \vect(\mb{Y}_1)\\
        \wb{\mb{z}} \triangleq \vect(\wb{\mb{Z}}) &= \mb{K}_2 \: \left( \mb{I}_N \otimes \mb{H}_2 \right) \: \vect(\mb{Y}_1)
    \end{align} \label{eq:def_z_zbar}
\end{subequations}
where $\mb{K}_0 \triangleq \mb{V}_0^T \otimes \mb{U}_0^T$ and $\mb{K}_2 \triangleq \mb{V}_2^T \otimes \mb{U}_2^T$. Rearranging the above equation, we get
\begin{equation}
    \wb{\mb{z}} = \mb{K}_2 \: \left( \mb{I}_N \otimes \mb{H}_2 \right) \: \mb{K}_0^{\dagger} \: \mb{z} \label{eq:lyap_like_vect}
\end{equation}
Observe that the number of unknowns in (\ref{eq:lyap_like_vect}) only depends upon the dimension $D$, i.e. $D^2 - D$ unknown elements in $\mb{z}$ and $D(D-1)/2$ elements corresponding to rotation matrix $\mb{H}_2$. However, the number of equations in (\ref{eq:lyap_like_vect}) depends on both $D$ and $N$ and is given by $(N - D)D + D$. This proves useful in defining the number of nodes required to solve (\ref{eq:lyap_like_vect}) for any dimension $D$. 

Consider the case for $D = 2$ and let $\mb{u} \in \bb{R}^2$ denote the unknown off-diagonal elements of $\mb{Z}_1$. We further denote the unknowns in rotation matrix $\mb{H}_2$ as $\mb{h} = \begin{bmatrix}
    h_1 & h_2
\end{bmatrix}^T$ where $\mb{H}_2 = \begin{bmatrix}
    h_1 & -h_2\\
    h_2 & h_1
\end{bmatrix}$ with the constraint $h_1^2 + h_2^2  = 1$. We can then rewrite (\ref{eq:lyap_like_vect}) as
\begin{equation}
    \mb{S} \: \wb{\mb{z}} = \mb{W} \: \bs{\phi}(\mb{u}, \mb{h}) \label{eq:basis_fcn}
\end{equation} where the unknown parameters in $\mb{H}_2$ and $\mb{z}$ correspond to $\mb{u}$ and $\mb{h}$, $\mb{S}$ is an appropriate selection matrix corresponding to the known elements of $\wb{\mb{z}}$. Here, $\bs{\phi}$ is a column of linearly independent scalar basis functions parameterized by unknowns $\mb{u}$ and $\mb{h}$ and $\mb{W}$ contains the corresponding coefficients. The problem is uniquely solvable if $\mb{W}$ is invertible, which is true for the given case since $\wh{\mb{B}}_1$ and $\wh{\mb{B}}_3$ are typically non-singular. For the set of basis functions in (\ref{eq:basis_fcn}), uniqueness of $\bs{\phi}(\mb{u}, \mb{h})$ also implies uniqueness in its arguments. For $D=2$, the basis function in (\ref{eq:basis_fcn}) is given by 
\begin{equation}
    \bs{\phi}(\mb{u}, \mb{h}) = \begin{bmatrix}
h_1 & h_2 & h_1 \: u_1 & h_1 \: u_2 & h_2 \: u_1 & h_2 \: u_2
\end{bmatrix}^T \label{eq:unique_basis}
\end{equation}
The solution to (\ref{eq:basis_fcn}) gives a unique set of basis function, $\wh{\bs{\phi}}(\mb{u}, \mb{h})$. For the given set of basis function in (\ref{eq:unique_basis}), the unique arguments $\wh{\mb{u}}$ and $\wh{\mb{h}}$ can be calculated as
$$\wh{h}_1 = h_1; \quad \wh{h}_2 = h_2; \quad \wh{u}_1 = \frac{h_1 \: u_1}{\wh{h}_1}; \quad \wh{u}_2 = \frac{h_2 \: u_2}{\wh{h}_2}$$
Hence, uniqueness in $\bs{\phi}(\mb{u}, \mb{h})$ implies uniqueness in its arguments, $\mb{u}$ and $\mb{h}$. With the estimate $\wh{\mb{u}}$, corresponding to the unknown elements of $\mb{z}$, $\wh{\mb{Y}}_1$ can be estimated using the relation in (\ref{eq:def_z_zbar}). Thus, we have the estimates of relative velocity $\wh{\mb{Y}}_1$, together with the estimates of relative position, $\wh{\mb{Y}}_0$, and relative acceleration, $\wh{\mb{Y}}_2$, from (\ref{eq:sol_const_acc_1}) at $t = 0$. The aforementioned steps involved in estimating the relative kinematics is summarised in Algorithm \ref{tab:algo}. 

\begin{algorithm}
\caption{Relative kinematics without accelerometer}\label{tab:algo}
\begin{algorithmic}[1]
\State \textbf{Input:} EDMs, $\mb{D}_k$ for all $t_k$, $k \in \{0, \ldots, K\}$.
\State For all $t_k$, evaluate the Grammian $\mb{G}_k$ using (\ref{eq:centered_grammian}).
\State Estimate $\mb{B}_l$ from (\ref{eq:grammian_coeff_est_no_acc}).
\State Estimate $\wh{\mb{Y}}_0$ and $\wh{\mb{Y}}_2$ from (\ref{eq:sol_const_acc_1}).
\State Estimate $\wh{\mb{Y}}_1$ and rotation $\mb{H}_2$ using (\ref{eq:lyap_cnst_acc_1}).
\State \textbf{Output:} $\wh{\mb{Y}}_0$, $\wh{\mb{Y}}_1$, $\wh{\mb{Y}}_2$ and $\mb{H}_2$.
\end{algorithmic}
\end{algorithm}


\section{Pairwise Distances and Accelerometer} \label{sec:data_model_acc}

We now consider a scenario where all the nodes have an accelerometer, and subsequently extend our existing data model to incorporate these accelerometer measurements. 
In the first step, we estimate the polynomial coefficients $\wt{\mb{Y}}_l$ for $l \geq 2$ in (\ref{eq:poly_traj_centered}) using the accelerometer measurements as given by (\ref{eq:acc_default}). In the second step, we use the estimates from the first step to modify the data model from (\ref{eq:time_poly_gram_vec}).

\subsection{Accelerometer measurement model} \label{sec:accelerometer}
The accelerometer measurement model for mobile node $i$ at time $t_k$, is given by
\begin{equation}
    \ddot{\wt{\mb{x}}}_{i,k} = \mb{Q}_{i,k} \: \ddot{\mb{x}}_{i,k}  + \epsilon_{a, k} \label{eq:acc_default}
\end{equation}
where $\ddot{\wt{\mb{x}}}_{i,k}, \ddot{\mb{x}}_{i,k} \in \bb{R}^{D}$ are the noisy and true acceleration (centered at the origin) for node $i$ at time $t_k$ and $\mb{Q}_{i,k}$ is the corresponding rotation matrix associated with the accelerometer at node $i$. The measurements are accompanied by white Gaussian noise i.e., $\epsilon_a \sim \mc{N}(0, \sigma^2_a)$ \cite[Chapter 2]{Manon2017}. Without the loss of generality, we assumed a calibrated accelerometer.

\textit{Assumption}: The data model for fusing the accelerometer measurements is proposed under the assumption that the mobile nodes are non-rotating. In other words, the accelerometer readings are measured w.r.t. a non-rotating frame of reference i.e., $\mb{Q}_{i,k} = \mb{Q} \in \bb{R}^{D \times D}, \: \forall t_k$. This is a feasible assumption for holonomic motion systems. The proposed data model can be extended to the cases where the orientation of individual mobile node is distinct and unknown but constant.

Stacking all the accelerometer measurements from  all the $N$ nodes we have
\begin{equation}
    \ddot{\wt{\mb{X}}}_k = \mb{Q} \: \ddot{\mb{X}}_k +  \mb{E}_{a, k} \label{eq:acc_default_mat}
\end{equation}
where the $i^{th}$ column of $\ddot{\wt{\mb{X}}}_k \in \bb{R}^{D \times N}$ corresponds to the accelerometer measurement from node $i$ at time $t_k$, $\ddot{\mb{X}}_k$ is given by (\ref{eq:poly_traj_centered}), and $\mb{E}_{a, k}$ represents the stochastic error.

\subsection{Coefficient Estimates from Accelerometer}
\begin{algorithm}
\caption{Relative kinematics with accelerometer}\label{tab:algo2}
\begin{algorithmic}[1]
\State \textbf{Input:} $\mb{D}_k$ and $\ddot{\wt{\mb{X}}}_k$ for all $t_k$, $k \in \{0, \ldots, K\}$.
\State Estimate $\wt{\mb{Y}}_l$ for $l \geq 2$ using (\ref{eq:acc_est}).
\State For all $t_k$, evaluate $\wt{\mb{G}}_k$ using (\ref{eq:gramian_coeffs_acc}).
\State Estimate $\wt{\mb{B}}_l$ from (\ref{eq:grammian_coeff_est_acc}).
\State Estimate $\wh{\mb{Y}}_0$ as given in (\ref{eq:sol_const_acc_1a}).
\State Estimate $\wh{\mb{Y}}_1$ and rotation $\mb{Q}$ using (\ref{eq:lyap_cnst_acc}).
\State Evaluate $\wh{\mb{Y}}_l = \mb{Q} \: \wt{\mb{Y}}_l$ for $l \geq 2$.
\State \textbf{Output:} $\wh{\mb{Y}}_0$, $\wh{\mb{Y}}_1$, $\wh{\mb{Y}}_2$ and $\mb{Q}$.
\end{algorithmic}
\end{algorithm}
\begin{figure*}[!t]
\begin{subequations}
    \begin{align}
    \mb{X} &= \begin{bmatrix}
    -244 & 385 & 81 & -19 & -792 & -554 & -965 & -985 & -49 & -503]\\
    -588 & -456 & -992 & -730 & 879 & 970 & 155 & 318 & -858 & 419
    \end{bmatrix}\\
    \mb{Y}_1 &= \begin{bmatrix}
    -5 & -8 & -6 & 6 & -1 & 2 & 1 & -5 & 9 & -5\\
    -8 & -5 & -7 & -9 & -3 & -2 & -2 & -10 & 2 & -1
    \end{bmatrix}\\
    \mb{Y}_2 &= \begin{bmatrix}
    -0.17 & -0.42 & 0.22 & -0.07 & 0.21 & -0.15 & 0.55 &  -0.72 & -0.49 & -0.34\\
    0.42 & 0.17 & 0.98 & 0.73 & 0.48 & 0.08 & -0.43 & -0.14 & 0.56 & 0.91
    \end{bmatrix} 
\end{align}\label{eq:true_coeffs}
\end{subequations}
    \hrulefill
    \vspace*{4pt}
\end{figure*}

    
Under the assumption of non-rotating reference frame for the accelerometers, the measurements for node $i$, using (\ref{eq:poly_traj_centered}), is given by
\begin{equation}
    \ddot{\wt{\mb{x}}}_k = \sum_{l = 2}^{L-2} \left( (l - 2) ! \right)^{-1} \: \wt{\mb{y}}_l \: t_k^{l - 2} + \epsilon_{a, k} \label{eq:acc_assumptions}
\end{equation}
where $\ddot{\wt{\mb{x}}}_k = \vect(\ddot{\wt{\mb{X}}}_k)$ and $\wt{\mb{y}}_l = \vect(\mb{Q} \: \mb{Y}_l)$ for $l \geq 2$. Stacking $K$ timestamps together in a column, we have
\begin{equation}
    \wc{\bs{\tau}} = \mb{V} \: \bs{\alpha}
\end{equation}
where $\mb{V} = \begin{bmatrix}
\mb{1}_K \otimes \mb{I}_{ND}, & \mb{t_k} \otimes \mb{I}_{ND} & \ldots & \mb{t_k}^{\odot L-2} \otimes \mb{I}_{ND}
\end{bmatrix}$, $\bs{\alpha} = \begin{bmatrix}
\wt{\mb{y}}_2, & \wt{\mb{y}}_3, & \ldots & \wt{\mb{y}}_L
\end{bmatrix}^T$, $\wc{\bs{\tau}} = \begin{bmatrix}
\wc{\bs{\tau}}_0, & \wc{\bs{\tau}}_1, & \ldots & \wc{\bs{\tau}}_K 
\end{bmatrix}^T$ with $\wc{\bs{\tau}}_k = \ddot{\wt{\mb{x}}}_k)$. The closed form estimate for the accelerometer coefficients can be obtained by solving the following least-squares problem $\arg \min_{\bs{\alpha}} \norm{\mb{V} \: \bs{\alpha} - \wc{\bs{\tau}}}_2^2$ leading to
\begin{equation}
    \wh{\bs{\alpha}} = \left(\mb{V}^T \: \mb{V}\right)^{-1} \: \mb{V}^T \: \wc{\bs{\tau}} \label{eq:acc_est}
\end{equation}
which is an optimal unbiased estimate of the acceleration coefficients, $\wt{\mb{y}}_l$, given the noise assumption.

\subsection{Data Model with Accelerometer Measurements}
Given estimates $\wt{\mb{Y}}_l$, $l\geq2$ are available from (\ref{eq:acc_est}), the formulation in (\ref{eq:time_poly_gramian}) can be modified such that
\begin{equation}
    \wt{\mb{G}}_k = \wt{\mb{B}}_0 + \wt{\mb{B}}_1 \: t_k + \wt{\mb{B}}_2 \: t_k^2 + \ldots + \wt{\mb{B}}_{L-1} \: t_k^{L-1} \label{eq:gramian_coeffs_acc}
\end{equation}
where $\wt{\mb{B}}_l = \sum_{m=0; m \neq l, \forall l > 2}^l \: \left( m! \: (l - m)! \right)^{-1} \: \wt{\mb{Y}}_m^T \: \wt{\mb{Y}}_{l-m}$ for $l \in \{ 0, 1, \ldots, L-1 \}$ and $\wt{\mb{G}}_k = \mb{G}_k - \sum_{l=2}^L \: \left( l! \right)^{-2} \: \wt{\mb{Y}}_l^T \: \wt{\mb{Y}}_l$. Here, we define $\wt{\mb{Y}}_l = \mb{Y}_l$ for $l \leq 1$. Vectorizing (\ref{eq:gramian_coeffs_acc}), we get
\begin{equation}
\wt{\mb{r}}_k = \wt{\mb{b}}_0 + t_k \: \wt{\mb{b}}_1 + t_k^2 \: \wt{\mb{b}}_2 + \ldots + t_k^L \: \wt{\mb{b}}_{L-1} \label{eq:gramian_coeffs_acc_vec}
\end{equation}
where $\wt{\mb{b}}_l = \vecth(\wt{\mb{B}}_l)$, for $l \in \{ 0, 1, \ldots, L-1 \}$ and $\wt{\mb{r}}_k = \vecth \left(\wt{\mb{G}}_k \right)$. Without loss of generality, let $\wc{\mb{r}}_k = \wt{\mb{r}}_k + \bs{\eta}_r$ be the noisy measurement plagued by additive white Gaussian noise with covariance matrix $\bs{\Sigma}_{r_k}$. Stacking all $K$ timestamps in column vector $\wc{\mb{r}}$, (\ref{eq:gramian_coeffs_acc_vec}) can be extended as,
\begin{equation}
    \wt{\mb{T}} \: \wt{\bs{\theta}} = \wc{\mb{r}} \label{eq:lsq_acc}
\end{equation}
where $\wt{\mb{T}} = \begin{bmatrix}
\mb{1}_K \otimes \mb{I}_{\wb{N}}, & \mb{t} \otimes \mb{I}_{\wb{N}}, & \ldots, & \mb{t}^{\odot L-1} \otimes \mb{I}_{\wb{N}}
\end{bmatrix}$, $\wt{\bs{\theta}} = \begin{bmatrix}
\wt{\mb{b}}_0, & \wt{\mb{b}}_1, & \ldots, & \wt{\mb{b}}_{L-1}
\end{bmatrix}^T$ and $\wc{\mb{r}} = \begin{bmatrix}
\wc{\mb{r}}_0, & \ldots, & \wc{\mb{r}}_K 
\end{bmatrix}^T$. Again, using the closed form solution for the least-squares problem $\arg \min_{\wt{\bs{\theta}}} \norm{\wt{\mb{T}} \: \wt{\bs{\theta}} - \wc{\bs{r}}}_2^2$, we have
\begin{equation}
    \wh{\wt{\bs{\theta}}} = \left( \wt{\mb{T}}^T \: \wt{\mb{T}} \right)^{-1} \: \wt{\mb{T}}^T \: \wc{\mb{r}} \label{eq:grammian_coeff_est_acc}
\end{equation} which again is an optimal estimator under additive white  Gaussian noise assumption on the measurements. The relative position estimate at time $t = 0$ can be calculated by solving for $\mb{Y}_0$ in (\ref{eq:sol_const_acc_1}a). As noted in (\ref{eq:acc_assumptions}), the estimate $\wt{\mb{Y}}_2$ from (\ref{eq:acc_est}) has an unknown rotation $\mb{Q}$ corresponding to the non-rotating accelerometer frame that needs to be estimated. Hence, to estimate the remaining unknowns, $\mb{Y}_1$ and $\mb{Q}$, consider the following set of equations
\begin{subequations}
    \begin{align}
        \wh{\mb{B}}_1 &= \wh{\mb{X}}_0^T \: \mb{Y}_1 + \mb{Y}_1^T \: \wh{\mb{X}}_0 \\
        2 \: \wt{\mb{B}}_3 &= \wh{\wt{\mb{Y}}}_2^T \: \mb{Q}^T  \: \mb{Y}_1 + \mb{Y}_1^T \: \mb{Q} \: \wh{\wt{\mb{Y}}}_2
    \end{align} \label{eq:lyap_cnst_acc}
\end{subequations}
which can be solved for $\mb{Y}_1$ and $\mb{Q}$ using the solving scheme introduced in section \ref{sec:no_acc_kin}. Algorithm \ref{tab:algo2} summarizes the intermediate steps as laid out in this section. 

\section{Simulation} \label{sec:sim}
\begin{figure}[htbp]
    \centering
    \includegraphics[width=0.55\textwidth,keepaspectratio]{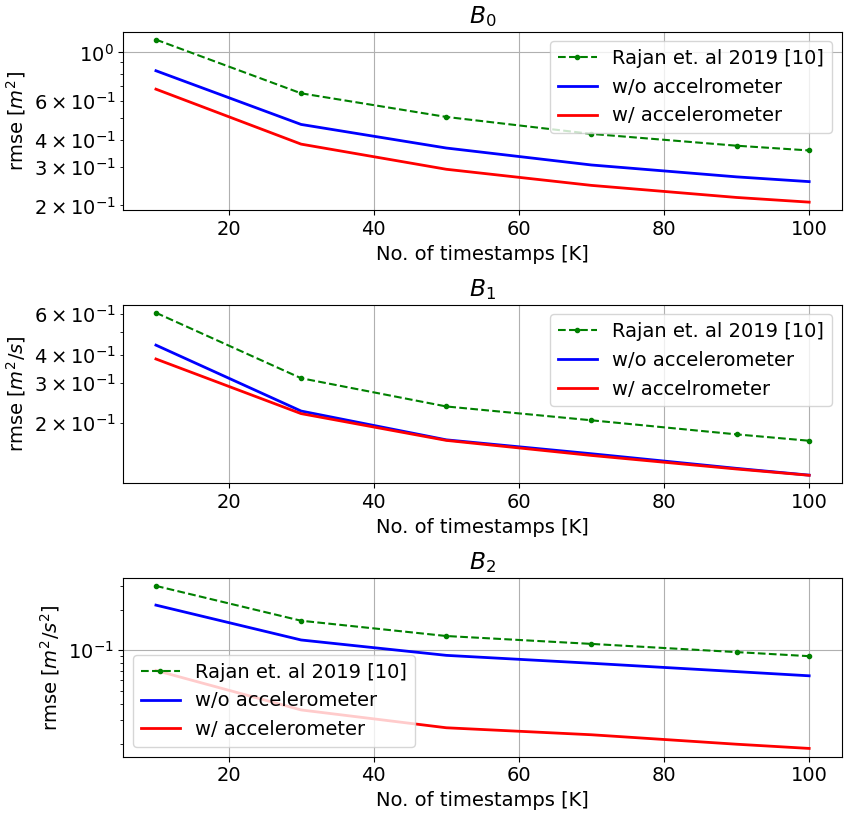}
    \captionsetup{width=\linewidth} 
    \caption{RMSE for the coefficient estimates $B_i$, $i \in \{0, 1, 2 \}$ in (\ref{eq:grammian_coeff_est_no_acc}) for the case without accelerometer and (\ref{eq:grammian_coeff_est_acc}) for the case with accelerometer, $\sigma_d = 0.01 \: m$ and $\sigma_a = 0.001 \: m/s^2$}        
    \label{fig:cnst_acc_coeff}
\end{figure}

\begin{figure}[!t]
    \begin{subfigure}[b]{0.42\textwidth}
        \centering
        \includegraphics[width=0.9\textwidth,keepaspectratio]{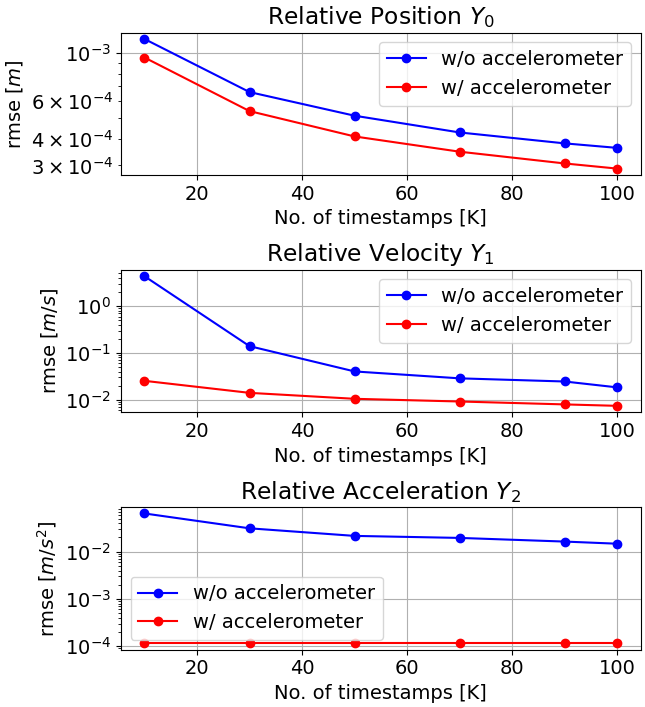}
        \captionsetup{width=0.9\linewidth} 
        \caption{RMSE on the relative kinematic estimates for varying $K$}
        \label{fig:cnst_acc_varying_K}
    \end{subfigure}
    \begin{subfigure}[b]{0.7\textwidth}
        \centering
        \hspace*{-10mm}
        \includegraphics[width=0.8\textwidth,keepaspectratio]{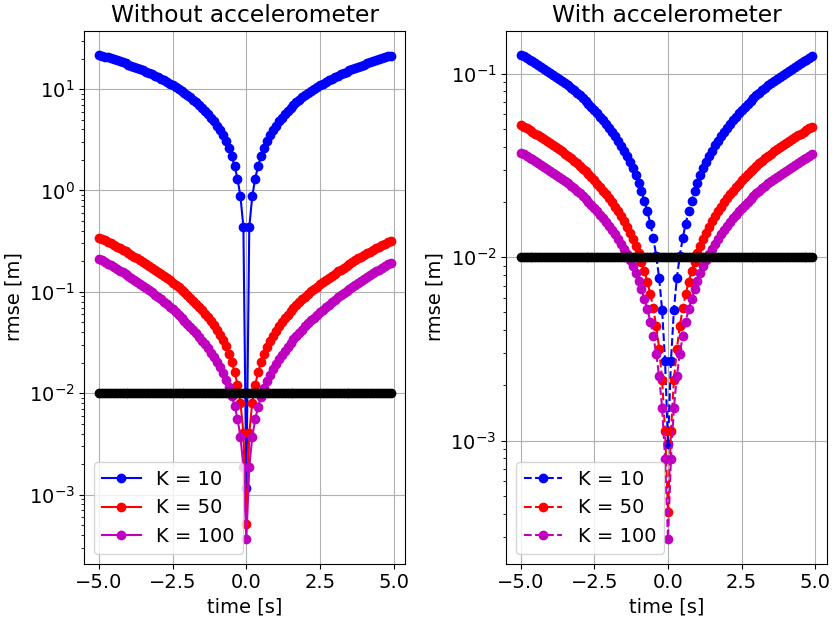}
        \captionsetup{width=0.7\linewidth} 
        \caption{RMSE on the time-varying positions under constant acceleration for varying $K$}
        \label{fig:cnst_acc_time_plots}
    \end{subfigure}
    \caption{a) Root-mean square error for relative position, relative velocity and relative acceleration at $t = 0$ for varying $K$. b) Root-mean square error for position measurements over time. For both plots, $\sigma_d = 0.01 \: m$ and $\sigma_a = 0.001 \: m/s^2$}
\end{figure}
\normalsize
For the simulation setup, consider a scenario with $N = 10$ mobile nodes in $D = 2$ dimensions, whose position, velocity and acceleration are given in (\ref{eq:true_coeffs}). The noise in the measurements, pairwise distance and accelerometer, are modelled as zero-mean Gaussian noise with a standard deviation of $\sigma_d = 0.01 \: m$ and $\sigma_a = 0.001 \: m/s^2$ respectively. A total of $N_{\text{exp}} = 1000$ Monte-Carlo runs were executed, and we compute the root mean square error for the parameters of interest as $\text{RMSE}(\mb{z}) = N_z^{-1} \: \left( \sqrt{N_{\text{exp}}^{-1} \: \sum_{i=1}^{N_{\text{exp}}} \norm{\wh{\mb{z}}_i - \mb{z}}^2} \right)$ where $\mb{z} \in \{ \wb{\mb{x}}, \wb{\mb{y}}_1, \wb{\mb{y}}_2 \} \in \bb{R}^{N_z}$. All the simulations are performed for a fixed time interval of $\Delta T = [-5, 5]$ seconds with varying values of $K$. 

Figure \ref{fig:cnst_acc_coeff} compares the estimates of the polynomial coefficients given in (\ref{eq:grammian_coeff_est_no_acc}) and (\ref{eq:grammian_coeff_est_acc}), for the case with and without acceleration respectively, w.r.t. the state-of-the-art in \cite{Rajan2019} (green curves). The proposed data model shows a lower root-mean square error (RMSE) for all the coefficient estimates when compared to \cite{Rajan2019}. Moreover, the addition of accelerometer measurements (red curves) lead to improvements in these estimates compared to the case when using only pairwise distances (blue curves). In addition to these improvements, the estimation of relative kinematics in \cite{Rajan2019} involving polynomial trajectories of order $2$ or more requires additional rigid-body constraints, which is not the case for our proposed approach, due to the solving scheme introduced in Section \ref{sec:data_model_no_acc}.

Figure \ref{fig:cnst_acc_varying_K} shows the RMSE for the estimates of the relative position, velocity and acceleration at time $t=0$ for varying $K$. The addition of accelerometer measurements shows significant improvement when compared to the estimates obtained only using pairwise distances. This improvement is also seen in Figure \ref{fig:cnst_acc_time_plots}, which shows the RMSE estimates of time-varying position measurements over time, which is estimated by substituting the estimated relative kinematics in (\ref{eq:poly_traj_centered}). The proposed solution is most accurate at $t = 0$ and worsens as we move away because the Taylor approximation gets worse as we move away from the location where the approximation holds.

\section{Conclusions}\label{sec:conclusion}
In this paper, we proposed an alternate formulation to the problem of estimating the relative kinematics given time-varying pairwise distances between mobile nodes. A solving scheme is proposed to uniquely obtain the relative kinematic estimates without the need of additional rigid-body constraints. We also introduce accelerometer measurements, under the assumption that the mobile nodes do not rotate and the motion is holonomic. Our proposed solution outperforms the state of the art, and the incorporation of accelerometer measurements considerably improves the relative kinematic estimates.

\bibliographystyle{IEEEtran}
\bibliography{./References}

\end{document}